\documentclass[12pt]{iopart} 
\usepackage{graphicx}
\usepackage{lineno}
\usepackage{comment}

\begin{document}

\title[]{Critical Current, Lengthwise Fluctuations, and Flux Jumps in REBCO CC: A Torque Magnetometry Study up to 45 T}
\author{J. Jaroszynski, A-M Constantinescu\footnote{Present address:
Department of Physics, University of New Mexico}, D. Kolb-Bond\footnote{Present address:
Commonwealth Fusion Systems}, A. Francis\footnote{Present address:
Commonwealth Fusion Systems}, A. Xu, R. Ries, G. Bradford\footnote{Present address: Advanced Conductor Technologies}, J. Bang, J. Lee, and D. Larbalestier}

\address{National High Magnetic Field Laboratory, Florida State University
Tallahassee, FL 32310, USA}
\ead{jaroszy@magnet.fsu.edu}
\vspace{10pt}
\begin{indented}
\item[] (\today)
\end{indented}

\begin{abstract}
REBCO coated conductors enable ultra-high-field superconducting magnets, but their exceptionally high critical currents make quantitative characterization at low temperatures and high magnetic fields challenging. In particular, conventional transport measurements and fixed-angle magnetometry fail near the \textit{ab}-plane, where the critical current is maximal.

We demonstrate a torque magnetometry method based on sample rotation in a constant magnetic field, enabling quantitative determination of the angular critical current $I_c(B,T,\theta)$ under conditions where other techniques are ineffective. Rotation generates a large variation of the perpendicular field component, allowing full current penetration and access to the critical-state magnetic moment. Measurements up to 45~T and down to 4.2~K are validated by numerical modeling.

The method reveals robust conductor performance up to 45~T, significant lengthwise $I_c$ variations—especially in edge-derived tapes—and thermomagnetic instabilities in the form of flux jumps. These results establish torque magnetometry as a practical tool for high-field characterization of REBCO conductors and provide insights relevant for magnet design.
\end{abstract}

\section{Introduction}
Rare-earth barium copper oxide (REBa$_2$Cu$_3$O$_{7-\delta}$, REBCO) coated conductors (CCs) have enabled major advances in high-field superconducting magnets, including all-superconducting magnets exceeding 30~T and hybrid systems approaching 50~T~\cite{32T,Pekin32,NATURE2019,Jonathan2026,Rossi}. These conductors are now central to the development of next-generation technologies such as high-field NMR, accelerator magnets, and fusion devices~\cite{fusion,Molodyk2023}. Their technological importance stems from their ability to sustain exceptionally high critical current densities $J_c$ at low temperatures and high magnetic fields.

Despite these advances, quantitative characterization of REBCO CCs under operating conditions remains challenging. In particular, the determination of the critical current $I_c(B,T,\theta)$, where $\theta$ is the angle between the magnetic field and the sample normal, is difficult at low temperatures and high magnetic fields. Under such conditions, $I_c$ often reaches kiloampere levels, making conventional transport measurements impractical due to Joule heating, limited current supply, and space constraints in high-field magnets. As a result, reliable data in the most relevant regimes---high field, low temperature, and angles near the \textit{ab}-plane---remain scarce.

Magnetometry-based approaches offer an alternative route by exploiting inductively generated screening currents. In these methods, changes in magnetic flux induce circulating currents in the conductor, producing a magnetic moment that can be related to $I_c$. However, standard magnetometry techniques, such as vibrating sample magnetometry, are not well suited for REBCO CCs in high-field environments due to large magnetic moments, strong torques, and electromagnetic noise in resistive magnets. Furthermore, commonly used measurement protocols based on magnetic field sweeps at fixed angle fail to establish a fully penetrated critical state when the magnetic field is nearly parallel to the \textit{ab}-plane. In this regime, the perpendicular component of the magnetic field becomes too small to drive current reversal across the entire sample, leading to a systematic underestimation of $I_c$.

In this work, we address these limitations using torque magnetometry combined with sample rotation in a constant magnetic field. By rotating the sample, a large variation of the perpendicular field component is generated even at fixed field magnitude, enabling full current penetration and access to the critical-state magnetic moment over a wide angular range. This approach allows quantitative determination of $I_c(B,T,\theta)$ under conditions where both transport measurements and conventional magnetometry are ineffective.

Building on our previous work~\cite{Jaroszynski2022,patent}, we present a detailed description of the instrument and its operating principles, and demonstrate its application to a broad set of REBCO conductors measured in magnetic fields up to 45~T and temperatures down to 4.2~K. The method is validated by comparison with transport data and numerical modeling, showing that the induced screening currents approach the critical current over a wide angular range.

Using this technique, we reveal several features of practical importance for high-field applications. First, we observe significant lengthwise variations in $I_c$ within individual tapes, particularly in samples cut from the edges of production tapes. These variations are most pronounced near the \textit{ab}-plane and can reach tens of percent, highlighting limitations of conventional characterization methods performed at higher temperatures. Second, we identify thermomagnetic instabilities manifested as flux jumps in certain conductors, particularly those with thick REBCO layers and limited stabilization. These observations provide insight into conductor design constraints for reliable magnet operation.

Overall, this work establishes torque magnetometry with sample rotation as a powerful and practical tool for high-field characterization of REBCO coated conductors, enabling access to regimes that are critical for magnet design but remain difficult to probe using existing techniques.

\begin{center}
\begin{figure}[hbt]
\vspace{-0.1in}
\hspace{-11pt}\hfill\includegraphics[width=.5\linewidth]{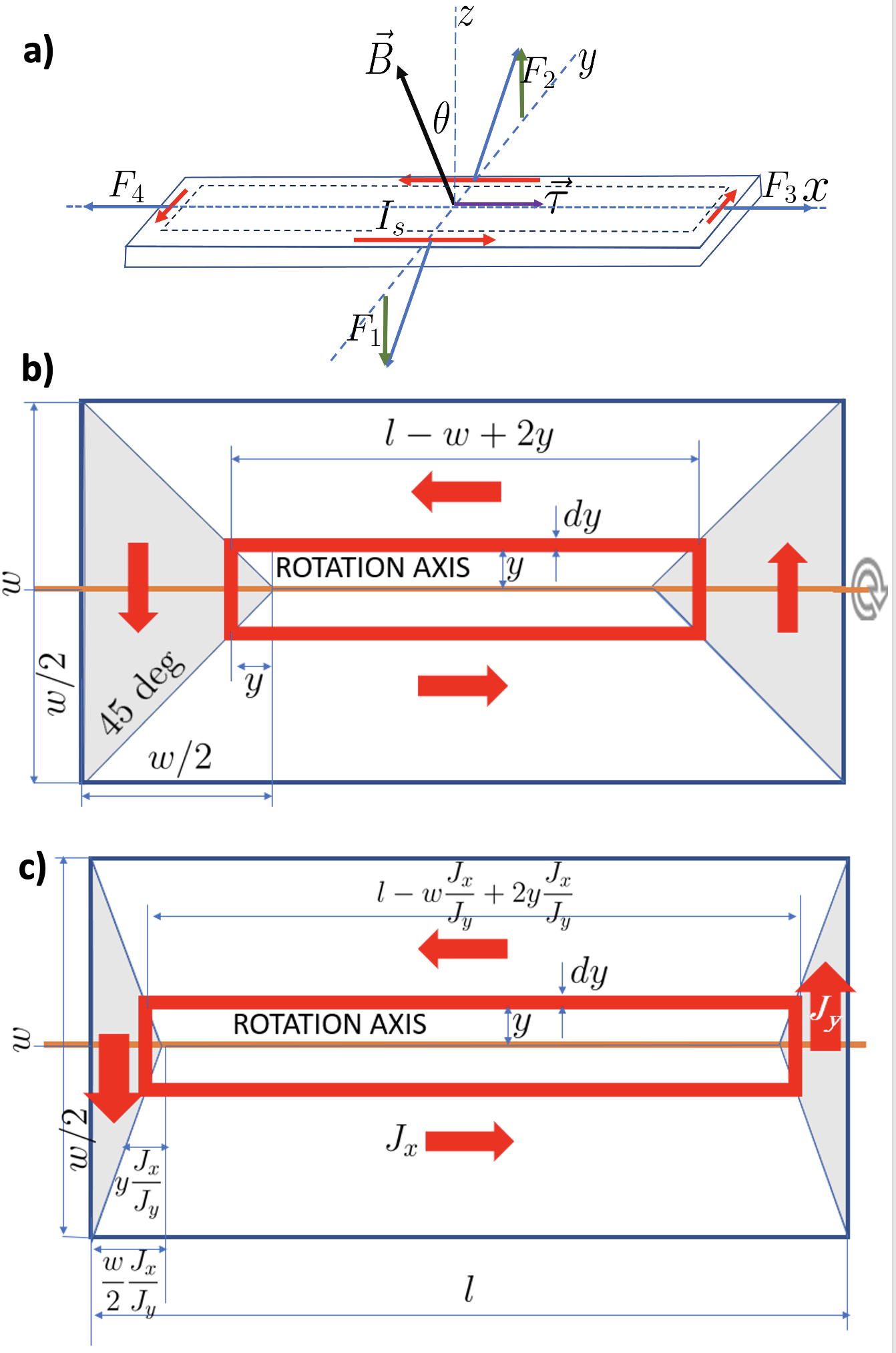}\hfill\hspace{1pt}\\
\vspace{-.3in}
  \caption[Gyorgy]{a) The forces exerted on a rectangular REBCO sample with an induced circulating screening current $I_s=I_c$, placed in a magnetic field $B$ tilted by an angle $\theta$ relative to the normal to the sample surface. This is an example when screening current is strong enough to prevent sample penetration by external magnetic field (still $\vec{B}$ is depicted in the sample center to show its direction only).  Only the vertical components of the Lorentz forces $F_1$ and $F_2$ contribute to the torque. The torque is parallel to $x$-axis. b) Distribution of screening currents in a homogeneous sample divided into loops to calculate the torque.  Here sample is fully penetrated by magnetic field.  c) Screening currents for a scenario where the current dg ensity across the sample, $J_c^x$, is greater than $J_c^y$ along its length. This situation arises when the sample is tilted in the magnetic field, and the current across the sample does not experience the full Lorentz force. The inset defines  $\theta$ as an angle between $\vec{B}$ and the normal to the sample plane. 
  \label{fig1}} \end{figure}
   \end{center}

\section{Torque and magnetic moment of screening currents}
Torque is the twisting force that causes an object to rotate around an axis. Torque is the cross product of the force and  the position vector $\vec{r}$ connecting the point at which the torque is  measured to the point where the force is applied.

$$\vec{\tau}=\vec{r}\times \vec{F}$$
$$\tau=rF\sin(\theta)$$

$\vec{\tau}$ is the torque vector and $\tau$ is the magnitude of the torque. $\vec{F}$ is the force vector, and $\theta$ is the angle between the force vector and  position vector.

Figure~\ref{fig1}a illustrates the torque exerted on a rectangular REBCO sample with an induced circulating screening current $I_s$, placed in a magnetic field $B$ tilted by an angle $\theta$ relative to the normal to the sample surface. This is the case when the magnetic field does not penetrate the sample. Screening currents flow only close to the edges and are sufficiently strong  to screen the field inside the sample. It is evident that only the vertical components of the Lorentz forces $F_1$ and $F_2$ contribute to the torque $\tau=(F_1+F_2)y\sin(\theta)$. The other components lie in the plane of the sample and cancel  each other, particularly the forces exerted on each end of the sample, where $F_3=-F_4$.
Thus, if $\vec{B}$ is perpendicular to the sample plane,  the torque vanishes. Figure~\ref{fig1}b shows the distribution of the screening currents depicted as the critical current density, $J_c$ in a homogeneous, rectangular sample. This is the situation when the sample is fully penetrated by the magnetic field, and screening currents flow in the entire sample. A discussion of how to reach such a saturated state is presented in  Section~\ref{sec3}. The torque can be calculated by dividing these currents into elemental loops

$$\tau=2BJ_ct\sin(\theta)\int_{0}^{w/2}dy(L-w+2y)y.$$

 After integration:

$$\tau=2BJ_ct\sin(\theta)\left(\frac{1}{2}ly^2|_0^{\frac{w}{2}}-\frac{1}{2}wy^2|_0^{\frac{w}{2}}+\frac{2}{3}y^3|_0^{\frac{w}{2}}\right)$$

$$\tau=2BJ_ct\sin(\theta)\left(\frac{1}{8}lw^2-\frac{1}{8}w^3+\frac{2}{24}w^3\right)$$

we get:

$$\tau=\frac{1}{4}BJ_ctw^2l\sin(\theta)\left(1-\frac{w}{3l}\right)$$

This can be rewritten as:

$$\vec{\tau}=\vec{m}\times\vec{B}$$
or

$$\tau=mB\sin(\theta)$$

where 

\begin{equation}
m=\frac{1}{4}J_ctw^2l\left(1-\frac{w}{3l}\right) \label{Gyorgy}
\end{equation}

where $\vec{m}$ is the magnetic moment, which characterizes the current distribution within the sample and determines its interaction with the external magnetic field. The above equation defines the magnetic moment.  This derivation was first proposed by Gyorgy et al.\ \cite{Gyorgy1989}.

Here, $l$, $w$, and $t$ are the film length, width, and thickness, respectively, and  $J_c$ is the critical current density. The sample rotates about an axis perpendicular to $B$ and parallel to its longer side $l$. This approach is valid when the critical current density $J_c$ is uniform. With $m$, we can obtain $I_c=J_cwt$ from Eq.~\ref{Gyorgy}.

Figure~\ref{fig1}c illustrates a similar integration approach for the case when $J_c^y > J_c^x$, that is, when $J_c^y$ (the current density across the sample) exceeds $J_c^x$ (the current density along the sample). In this case, the magnetic moment is expressed as:

\begin{equation}
m=\frac{1}{4}J_ctw^2l\left(1-\frac{w}{3l}\frac{J_c^x}{J_c^y}\right)
\end{equation}

This scenario arises when the sample is tilted around the rotation axis and the 'return' currents at either sample end shaded  in Fig.~\ref{fig1}  do not fully experience the Lorentz force, thereby artificially increasing the observed moment. This is because only the region where the current flows along the rotation axis contributes to the torque and this region becomes larger. In certain samples, $J_c^y<J_c^x$, notably in conductors grown on an inclined surface deposition (ISD) by THEVA \cite{Theva2004}. This, in turn, reduces the observed $m$. These effects are significantly mitigated if $l\gg w$.

Typically, $w=2-4$~mm and $l=10-15$~mm as these dimensions maintain the error below 10 \% even for $J_c^y\ll J_c^x$, as elaborated in \cite{Thompson2009}.

A more detailed and general derivation of the above expressions can be found in \cite{Rasto2026}.

\section{Design}

The magnetometer is specially engineered for  efficient
 assessment of the critical current in REBCO CC tapes as  a function of the field, field angle, and temperature. It is also useful for studying other materials with strong magnetic moments and anisotropy, such as Nd magnets.

This system features a centrally located rotating sample
 platform within an external magnetic field.  The platform is suspended by two Niva points anchored to a  yoke complemented
 by two V-shaped sapphire jewels (Swiss Jewels) at its extremities. This design enables an almost frictionless rotation.
A sample with dimensions of $4\times15$ mm$^2$ is positioned
 within a 100 $\mu$m deep groove parallel to the rotation axis. It is then
 covered with a flat piece and fastened to the platform  using 0-80 screws.
The sample platform, cover, and yoke are made from nonmagnetic polyamide (Meldin) or unalloyed titanium (grade 2).

 Parallel to the rotating platform, a pickup coil
 detects the angle between the sample and external magnetic
 field. For precise temperature control, the Cernox thermometer is pressed
 into the sample. A resistive heater (3W surface mount resistor)
 of size $1\times 1.5\times 3$ mm$^3$ is integrated onto the platform.

The  rotation of the sample is enabled by a 28 mm-diameter titanium pulley attached to the platform, linked with two short fishing lines that slightly surpass the  circumference of the spool. These lines are connected to two aluminum alloy (grade 6061) rods, each approximately 1.5 m in length and 3 mm in diameter. Aluminum was selected for its nonmagnetic properties because stainless steel was found to introduce significant forces owing to its inherent magnetism.

One rod connects to a load cell (e.g., WMC Load Cell 22N from Interface Inc.), which is situated at room temperature, roughly 1.6 m from the  center of the magnet. This cell is coupled with an actuator (from MDC Vacuum Products, Inc., Linear Motion FT, Standard, 4" Travel, Manual Drive) that offers a linear movement of 4 in. The actuator is driven by a servo-motor 
controlled by a computer. This mechanism allows the sample platform to rotate through more than 400~$^{\circ}$. However, during the typical characterization, the sample is rotated between 210~$^{\circ}$ and 240~$^{\circ}$.

A single  rotation of the actuator’s microscrew translates to a 1/40 inch linear movement, corresponding to a 2.72-degree rotation
 of the sample platform. A computer-controlled brushless servo motor drives the linear actuator at a wide range of speeds, from 0.01 to 2~$^{\circ}$/s.
 The second aluminum rod is connected to a  tensioning mechanism
 (a counterweight of $\sim700$~ G or a spring, $\sim200$~G/inch), which balances the rotating platform.

The sample holder is placed at the field center of either a 52 mm bore 15 T superconducting magnet or a 31 T resistive magnet (with a 39 mm inner diameter cold bore). The sample rig (with a 39 mm outer diameter) can be placed in a long vacuum can with a regulated amount of helium exchange gas, or directly in liquid helium (LHe). The outer diameter of the magnetometer yoke is 1.25” (31.75 mm).

\begin{center}
\begin{figure}[t]
\hspace{1pt}\hfill\includegraphics[width=.9\linewidth]{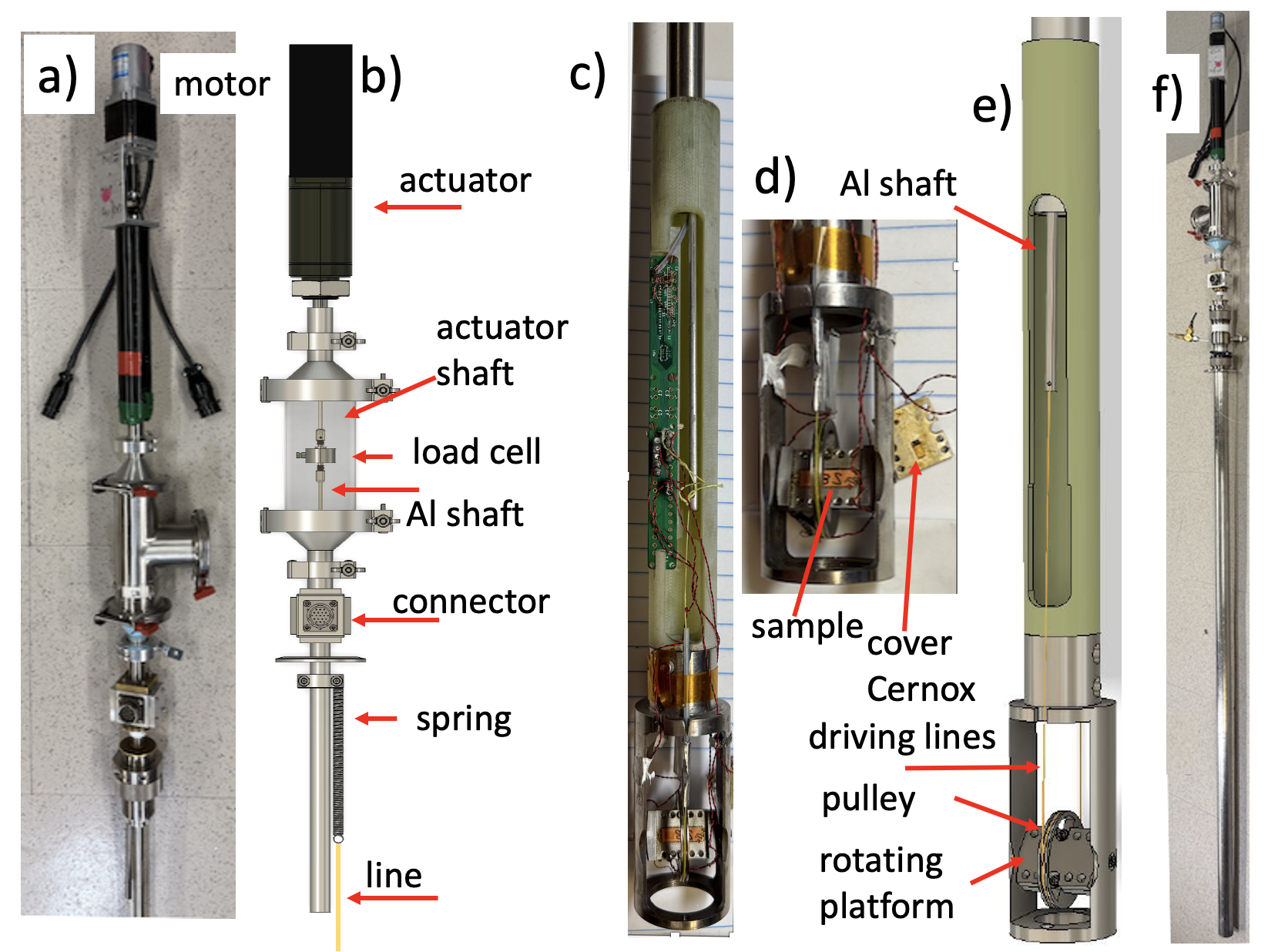}\hfill\hspace{1pt}\\
  \caption{Photographs (a,c,d,f) and drawings (b,e)  of the torque magnetometers used in resistive magnets. Description in the main text}\label{schemat}
\end{figure}
\end{center}

Measurements up to 45 T using the NHMFL hybrid magnet with a 16 mm bore are possible with a smaller probe. In such instances, the fishing lines are connected to a 10 mm diameter spool, leading to reduced precision in angle determination. Nevertheless, our apparatus remains compatible with the most powerful magnets and consistently ensures precise temperature control from 4.2 K to 50 K. The main features of the apparatus are  shown in
Figure~\ref{schemat}.

\section{Principle of operations\label{sec3}}
\subsection{Field sweeps at fixed angle}

Magnetometry is commonly performed by sweeping the magnetic field at a fixed angle between the sample and the applied field. In this configuration the screening currents in REBCO CC are generated by  the perpendicular component of the magnetic field. Representative hysteresis loops obtained by field sweeps at fixed angle are shown in Fig.~3a.  The  force derived from the loops width and plotted in Fig.~3b  agrees with the pinning force derived from the transport measurements \cite{Francis2020,Francis2026} at the same angle $\theta = 108^\circ$ (18$^\circ$ from the $ab$ plane) at different temperatures. The angular dependence of the torque is shown in Fig.~3c.

A notable feature of the torque profile obtained from field sweeps is the strong reduction of the signal near the $ab$ plane ($\theta = 90^\circ$). At this angle the torque vanishes even though the critical current $I_c(\theta)$ is known to reach a maximum. This behavior arises because the perpendicular field component
\begin{equation}
B_\perp = B \cos(\theta)
\end{equation}
approaches zero when the magnetic field becomes parallel to the $ab$ plane. In this regime the screening currents generated by the field sweep remain confined to the edges of the sample and do not penetrate the entire conductor. As a result, a fully developed critical state cannot be established.

Thus, even large field sweeps of $\pm 30$~T are insufficient to reverse currents across the full sample when $\theta \approx 90^\circ$. Indeed, it is well established \cite{Uglietti2009,Aixia2010} that $I_c(\theta)$ exhibits a  maximum around this angle. This causes the collapse of the hysteresis loops in Fig.~3a and the suppression of torque in Fig.~3c. Thus, the drop of torque near the \textit{ab}-plane reflects a not fully penetrated state. Consequently, torque measurements based on field sweeps underestimate the critical current when the magnetic field is nearly parallel to the $ab$ plane. 

Therefore, field sweeps at fixed angle do not provide reliable $I_c(\theta)$ near the \textit{ab}-plane, where $I_c$ is often maximal and most relevant for applications.

\begin{center}
\begin{figure*}[htb]
\hspace{1pt}\hfill\includegraphics[width=0.9\linewidth]{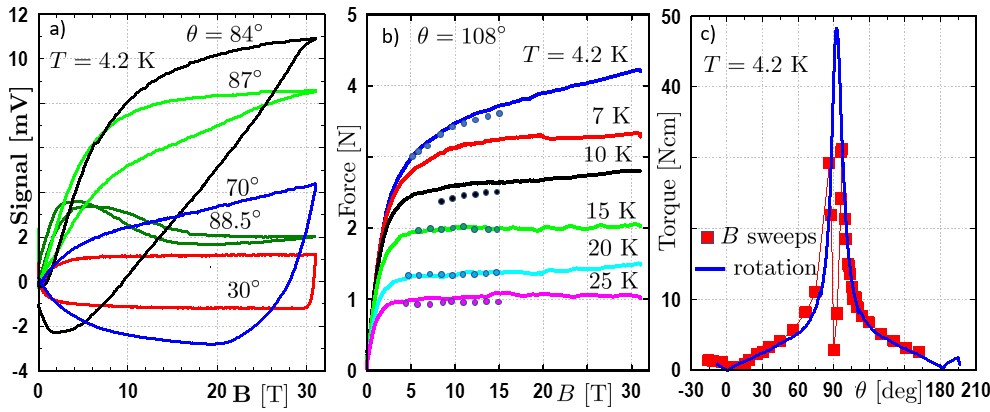}\hfill\hspace{1pt}\\
\caption{a) Load cell signal hysteresis loops   when the magnetic field is swept at various fixed angles. b) The force  assessed from the  hysteresis loops (solid lines) and transport (bullets) measured at $\theta=108$~deg, that is 18$^\circ$ from the ab plane at different temperatures. c) The torque as a function of angle at 4.2 K and 30~T. Square symbols represent results from field sweeps at fixed angles; the continuous line denotes result from the sample rotation at fixed field. Unlike the fixed angle measurements which drop near 90$^{\circ}$, the continuous angle sweep generate peaks near 90$^{\circ}$.}\label{fig3}
\end{figure*}
\end{center}

\subsection{Rotation at fixed magnetic field method}
To overcome this limitation, we rotate the sample in a constant magnetic field, thereby generating a large effective sweep of $B_\perp$,  enabling $I_c(\theta)$  measurements near the ab plane where conventional field sweeps are ineffective. 

During rotation, the flux varies as $\Phi=BS\cos(\theta) $,  producing a change in $B_\perp$ sufficient to drive full current penetration even near the ab-plane.  This assumption is verified numerically in Section 4.3. Once full penetration is reached, the screening currents flow throughout the sample at approximately $\pm I_c$, and the magnetic moment approaches the critical-state value described by Eqs.~(1) and (2). The torque then directly reflects $I_c(\theta)$ through Eq.~(1).

\begin{center}
\begin{figure}[h]
\hspace{1pt}\hfill\includegraphics[width=0.6\linewidth]{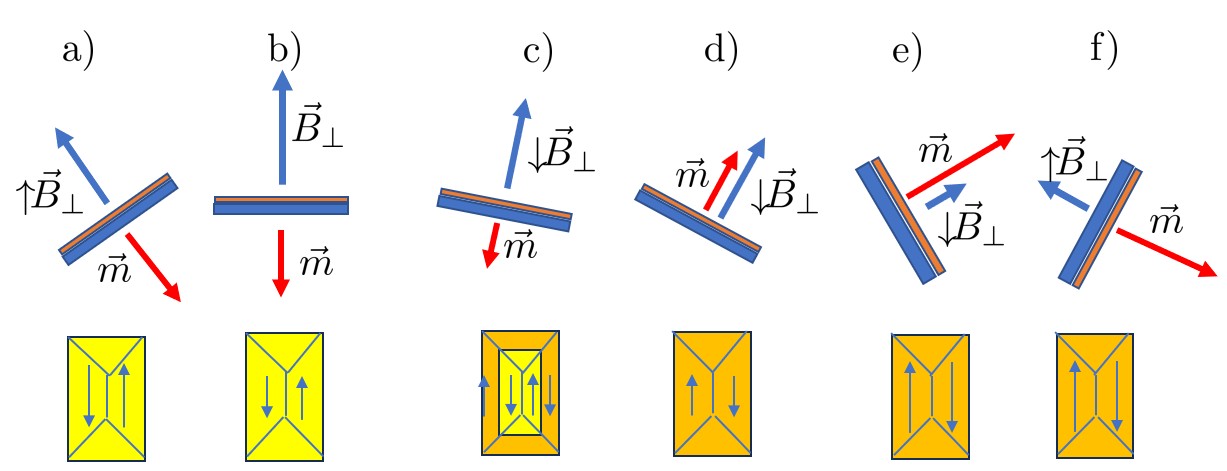}\hfill\hspace{1pt}\\
\caption{a) sample rotates clockwise toward $B\parallel\mbox{c-axis}$ configuration. Since the perpendicular field component $B_{\perp}$ increases, the moment from screening currents $\vec{m}$ has opposite direction to $\vec{B_{\perp}}$  to diminish magnetic flux changes b) sample is perpendicular to $\vec{B}$, the moment is still opposite. c) $B_{\perp}$ starts decreasing, the moment (and screening currents) start to reverse direction  but this takes a few degrees, depending on $I_c$. d)   $B_{\perp}$ decreases so the moment has the same direction. e) still $\vec{m}\parallel\vec{B_{\perp}}$the moment increases because $I_c$ is larger closer to the ab plane. f) after $\vec{B}$  rotates through the ab plane,
 $B_{\perp}$ starts increasing but at the same time $\vec{m}$ becomes opposite to $B_{\perp}$ just because sample changes orientation, without  the screening currents  reversal.}\label{fig4}
\end{figure}
\end{center}

During rotation, as shown in Fig.~\ref{fig4}, the reversal of the screening currents occurs only in a narrow angular interval after the field passes $\theta = 0^\circ$ and its perpendicular component starts to decrease. In contrast, when the field crosses the \textit{ab}-plane, the  currents do not reverse in the sample frame. This is the central advantage of the rotation method: it enables reliable determination of $I_c(\theta)$ in the angular region where field sweeps fail.

In Fig.~\ref{fig3}c, the magnetic moment of the sample, measured during rotation at $B=30$~T (represented by the solid line), is displayed alongside the data derived from field sweeps (denoted by symbols). Indeed, the rotation data exhibits a pronounced peak around the ab plane, which aligns with expectations and the above considerations.

The electric field generated by rotation is given by Eq.~\ref{eq4}, and the corresponding induced current follows Eq.~\ref{eq5}. Because of the large $n$ values typical of REBCO, the induced current depends only weakly on rotation speed and remains close to the critical current over a broad range of conditions.
\begin{equation} E=-\frac{d\Phi}{dt}/(2w+2l)=\omega BS\sin(\omega t)/(2w+2l)
\label{eq4} \end{equation}
where $t$ is time and typical $2w+2l=36$~mm is the sample circumference. For a rotation of one degree per second ($\omega\approx0.017$~rad/s), when the field $B=10$~T rotates through the ab plane, this yields approximately 2.7~$\mu$V/cm at the edge of the sample. This value is  similar to the electric field criteria used in the transport measurements. It changes with $\omega$, $B$, and the angle. However, for typical $n\approx30-50$ values, the induced current: 
\begin{equation} I\propto E^{1/n} \label{eq5}\end{equation}
barely depends  on these parameters.  This is another fortunate feature that makes the method useful and an appropriate substitute for the transport current measurements preferred for magnet qualification and design.

\subsection{Numerical validation}
Numerical modeling shows that screening currents exceed 98 \% of $I_c$ over a wide angular range, confirming the validity of the method.
We numerically evaluated how closely the screening currents approach saturation, that is to $I_c$ values. Calculations utilize  2D  $T$--$A$  model similar to that developed for axisymmetric
solenoids \cite{EBJ2019}. The effects of rotation have previously been investigated in the context of screening-current-induced fields (SCIF), mechanical strain, and changes in mutual inductance. \cite{Dylan2021,Yan2022,Noguchi2023}. 
We included  rotation by changing the surrounding 
background magnetic field.

The critical current is parametrized by  the Kim fit:
\begin{equation}
 J_c(B,\theta) = \frac{J_{c0}}{1+\left(\frac{\sqrt{k^2\sin^2(\theta)+\cos^2(\theta)}}{B_0/B}\right)^\alpha} 
\end{equation}
where     $J_{c0}= 2000$~A/cm$^2$, 
 $k=0.0630$, $B_0=1.56$~T, and 
 $\alpha= 1.088$,
and we include the power law of the 
superconducting resistivity:
\begin{equation}
 \rho = \frac{E_c}{J_c}\left(\frac{J}{J_c}\right)^{n}.
\end{equation}
with  $E_c = 1$~$\mu$V/cm and $n = 50$.  REBCO film thickness $ t_f = 1 ~\mu \mbox{m}$,  total tape thickness 
$ t_t = 100$~$  \mu \mbox{m}$, and  tape width
$ w = 4$~$ \mbox{mm} $. 

\begin{center}
\begin{figure}[htb]
\hspace{1pt}\hfill\includegraphics[width=0.5\linewidth]{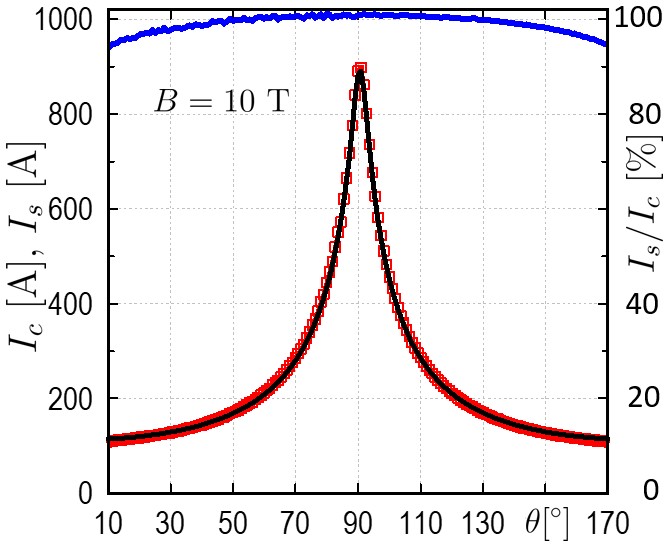}\hfill\hspace{1pt}\\
\caption{Calculated screening current (red symbols), critical current (black line) and their ratio (blue line) when sample rotates starting at $\theta=-30^{\circ}$ while field is set $B=10$~T}
\label{fig5}
\end{figure}
\end{center}

In Fig.\ref{fig5} we plot the screening currents, critical current and saturation as 
 $I_{sc} = \frac{1}{2}\int|J| dS $,
$I_{c} = \frac{1}{2}\int|J_c| dS$ ,
and  $f_{s} =  \int|\frac{J}{J_c}| dS$,
where the screening current is calculated for half of the sample (hence 1/2).

Figure~\ref{fig5} shows the calculated screening current, critical current, and their ratio for a sample rotated at fixed field. Over the angular range $30^\circ < \theta < 150^\circ$, the screening current exceeds 98\% of $I_c$ and it flows  at approximately  $\pm I_c$ throughout the sample.  This confirms that the magnetic moment measured during rotation is very close to its critical-state value,  validating the use of torque magnetometry for quantitative determination of $I_c(\theta)$ over the angular range of practical interest.

\begin{center}
\begin{figure}[h!]
\hspace{1pt}\hfill\includegraphics[width=0.99\linewidth]{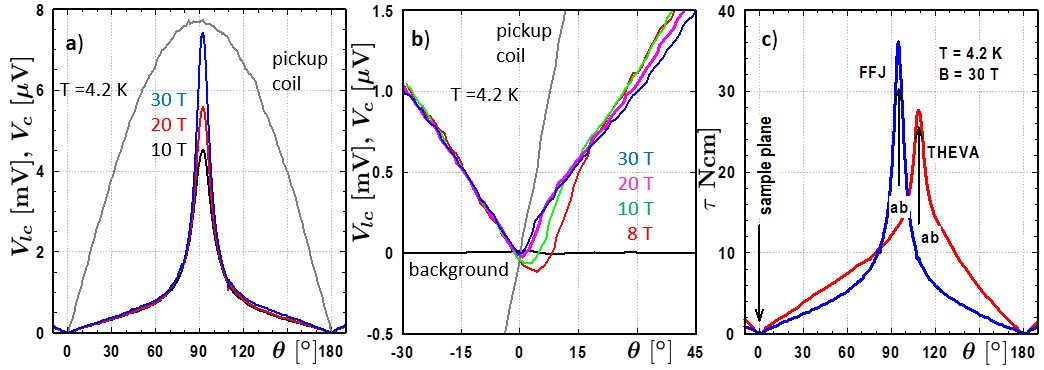}\hfill\hspace{1pt}\\
\caption{Load cell and pickup coil signals.
(a) Signals during sample rotation in three magnetic fields. The coil signal vanishes for $B$ perpendicular to the sample plane, where also $\tau=0$.
(b) Signals near $\theta=0^\circ$ for different fields. The background is the empty magnetometer at 30~T. At low fields, a small negative torque appears due to the high $I_c$ of the HM sample, in contrast to the low-current AP sample (left).
(c) Angular dependence of torque $\tau(\theta)$. Zeros occur at $\theta\approx0^\circ$ and $180^\circ$. The \textit{ab}-plane maxima are shifted by $\sim4^\circ$ (Faraday Factory) and $\sim18^\circ$ (Theva), reflecting the intrinsic tilt of \textit{ab}-planes from IBAD and  ISD growth, respectively.}
\label{fig6}
\end{figure}
\end{center}

\subsection{Torque measurements when sample rotates}
Figure~6a shows the signal from load cell (the force is proportional to it)  measured during rotation at fixed fields. The total signal contains contributions from the sample torque, the line-tension force, and residual friction. After subtraction of the background, the torque is obtained as $\tau = Fr$, where $r$ is the pulley radius.

A major practical advantage of the rotation method is that the torque baseline can be determined each time the field crosses the sample normal, that is, at $\theta \approx 0^\circ$ and $180^\circ$, where $\tau = 0$ by symmetry because the field is perpendicular to the sample plane. This allows accurate background subtraction during a single angular sweep.

 Figure~\ref{fig6}b shows that for samples with very high current near the c-axis, the reversal of the screening  currents can be observed. For $B=8$~T $I_c=1400$~A for this SuperPower HM sample. That high current needs rotation to $\theta\approx15^{\circ}$ for full reversal. In narrow range of angles the torque is negative (see Fig.~\ref{fig4}c). However at $B=30$~T the screening current reversal becomes unnoticed: the minimum of the torque is zero at exactly where the pickup coil determines sample is perpendicular to the field. This is because of lower  $I_c=400$~A and much higher field.

The torque technique is also useful in determination of the ab plane offset in REBCO CC.   REBCO tapes fabricated by  Ion-Beam Assisted Deposition have the geometric and crystallographic axes misaligned \cite{Maiorov2005}. The other example is  REBCO grown on  Inclined Substrate Deposition by Theva \cite{Theva2004}. In such samples the ab plane is tilted with regard to sample base. As a result the maximum of $I_c$ occurs not when $B$ is parallel to the sample plane ($\theta=90^{\circ}$), but when $B$ is parallel to the ab plane. However, it must be stressed that the currents, both transport and screening flow always in the sample plane. This is because REBCO plane is not tilted as a whole but rather grows as fish scales structure. Vertical roughness is of the order of micrometers that is negligible for millimeter sample dimensions. Because of that, the torque vanishes when sample, not crystallographic planes are perpendicular to the field. Angles where the torque is zero allow to determine sample orientation in addition to Hall probe or pickup coil. The ab peak is observed with an offset from $\theta=90^{\circ}$, usually a few degrees \cite{Jun2023,Jessica2025,Jonathan2025}, but substantial in Theva production \cite{Theva2004}, as it is shown in Fig.~\ref{fig6}c. 

Rotation thus provides both reliable $I_c(\theta)$ over the important angular range and a direct means to determine the offset of the \textit{ab}-plane from the sample surface.

In summary, field sweeps fail near the \textit{ab}-plane because the small perpendicular field component cannot establish the full critical-state magnetic moment. Rotation at fixed field overcomes this limitation by generating a large change in $B_\perp$, while the torque remains governed by Eq.~(1). The torque vanishes when the field is perpendicular to the sample plane, whereas the torque maximum tracks the orientation of the \textit{ab}-plane through the angular dependence of $I_c$.

\section{Experimental results}
\subsection{Critical current assessment up to 45~T.}
\begin{center}
\begin{figure}[htb]
\hspace{1pt}\hfill\includegraphics[width=0.8\linewidth]{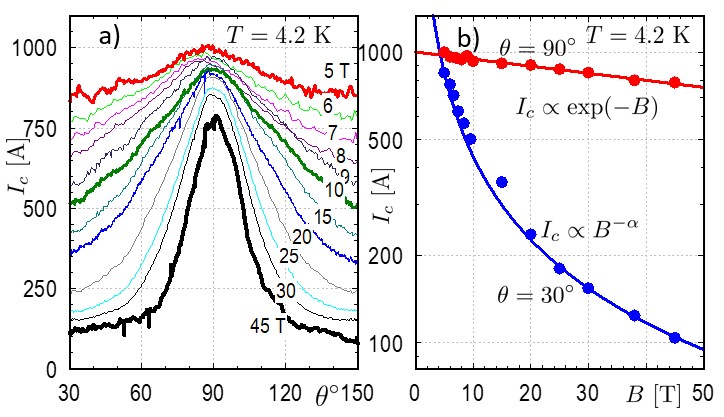}\hfill\hspace{1pt}\\
\caption{a)  Angular $I_c$ assessed using torque in R\&D sample  from SuperPower with 15~\% BZO, measured at $T=4.2$~K within a magnetic field range of 5 to 45 T. b) Critical current as a function of field for the field in the ab plane  and at $\theta=60$~deg off the ab plane.}
\label{fig7}
\end{figure}
\end{center}

Figure~\ref{fig7}a shows the angular critical current assessed in a sample at several fields ranging from 5~T  to 45~T.  Because of the small diameter of the space (16 mm), measurements in the 45~T hybrid magnet are made using a different probe with a smaller rotator pulley. This  increases the noise and  angle reading error. Figure~\ref{fig7}b shows $I_c$ measured at the ab plane and the c axis. The solid lines show the fits to the exponential and power functions, respectively. It is clearly seen that at $T=4.2$~K there is no  obvious  change in behavior  at high fields up to 45~T. These findings suggest that REBCO tapes maintain a strong performance even under extreme magnetic fields, which is crucial for their use in future  applications.

\subsection{Characterization of modern REBCO CC tapes.}

\begin{center}
\begin{figure*}[htb]
\hspace{1pt}\hfill\includegraphics[width=.99\linewidth]{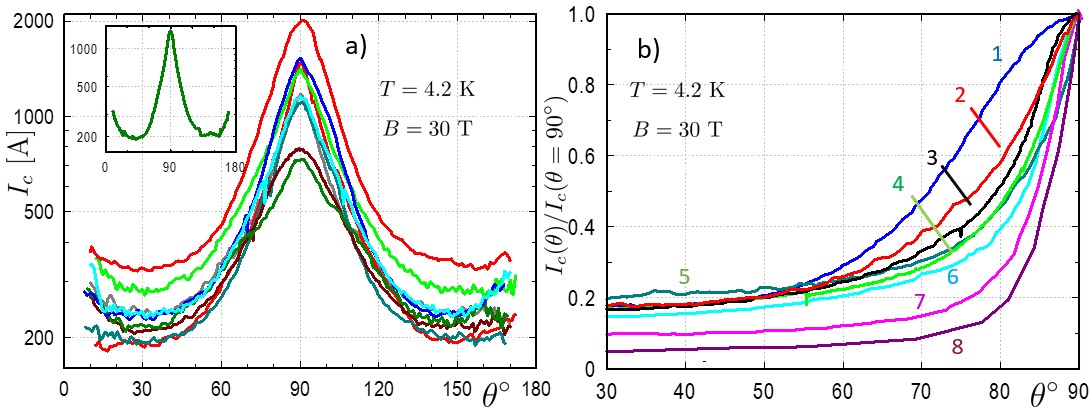}\hfill\hspace{1pt}\\
\caption{a) Angular $I_c$ assessed using torque in multiple samples of HM formulation with 15 \% BZO doping from SuperPower, measured at $T=4.2$~K at $B=30$~T  b) Angular critical current at $T=4.2$~K at $B=30$~T measured in samples from different manufacturers  normalized to  at the ab plane. 1- 15~\% BZrO$_3$ R\&D SuperPower, 2-15 \% BZrO$_3$ production SuperPower, 3- Eu doped SuperMag production, 4- Eu doped Shanghai SST production, 5- Y$_3$O$_3$ doped Faraday Factory production, 6- Y$_3$O$_3$ doped Faraday Factory production 'Mirai', 7- 7.5 \% BZrO$_3$ production SuperPower, 8- undoped SuperPower }\label{fig8}
\end{figure*}
\end{center}

Figure~\ref{fig8}a shows $I_c(\theta)$ assessed with torque in several production samples with 15 \% BZO from SuperPower Inc.\ at $B=30$~T and $T=4.2$~K.
 The samples have quite different magnitudes of $I_c$, ranging from 700 to 2000~A at the ab plane. This variability  makes it possible to select an appropriate  conductor for a given application. In addition, all conductors have  wide ab plane maxima with a full-width at half-maximum (FWHM) of approximately $20^{\circ}$. These maxima are accompanied by shallow minima at $\theta\approx 30^{\circ}$ and $\theta\approx 150^{\circ}$. These minima are a consequence of the weak and wide maxima around the c axis. It is well known that in BZO -doped samples,  strong pinning to BZrO$_3$ (BZO) nanorods  dominates at high temperatures, leading to a pronounced  maximum at the c axis. Because of the high BZO content, they are still observed at low temperatures. This leads to a relatively low anisotropy, which is beneficial for both magnet and cable applications.  Figure~\ref{fig8}b compares the REBCO tapes of different formulations and manufacturers. Significant differences between the ab plane maxima are clearly seen, from $\mbox{FWHM}=5^{\circ}$ for non-BZO formulation to $\mbox{FWHM}=32^{\circ}$ in the R\&D 15 \% sample. It is also seen that in non-BZO samples, $I_c$ decreases all the way from the ab-plane to the c-axis while it is flat beyond $\sim 30^{\circ}$ and $\sim 150^{\circ}$ at 7.5~\% doping and increases there for 15~\% doping. In general, the above results show that modern, heavily doped samples show very broad ab $I_c$ peaks and much lower anisotropy, both of which are beneficial for magnet construction.

\subsection{Lengthwise $I_c$ fluctuations}
REBCO CC are known to exhibit non-uniformity along their length and width. The lengthwise variation in the critical current \( I_c(x) \), where \( x \) is the position on the tape, can be measured relatively easily at 77 K using transport and remnant magnetization techniques \cite{PaulIEEE,lidia2016}. Commercial devices, such as Tapestar, are available for assessing \( I_c(x) \). Over the years, the standard deviation of \( I_c \) relative changes has decreased from approximately 10\% to 2-3\% in tapes from certain manufacturers. For older conductors, the variations were mainly due to changes in the cross-section of the REBCO film. In newer conductors with more stable cross-sections, variations in the pinning landscape along the tape dominate $I_c(x)$. These  originate from   an unstable production process, such as fluctuations in temperature during deposition. In practice any production step can contribute to such inhomogeneities. 

\begin{center}
\begin{figure}[bth]
\hspace{1pt}\hfill\includegraphics[width=0.8\linewidth]{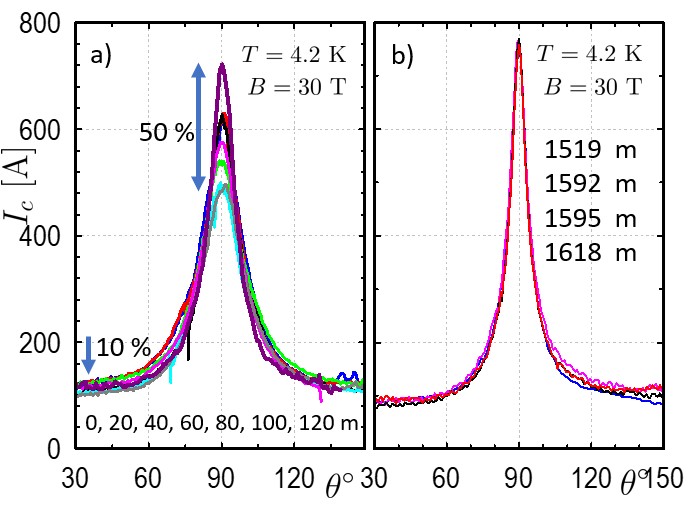}\hfill\hspace{1pt}\\
\caption{a) Angular $I_c$ for several samples cut from the edge slit of a long tape every 20 m  b) Angular critical current for four samples cut from the central slit of an another tape at positions between 1519 to 1618 m.  $T=4.2$K at $B=30$~T }\label{fig9}
\end{figure}
\end{center}

Characterization at higher temperatures (65-77 K) cannot be  accurately extrapolated to lower temperatures owing to the different pinning mechanisms that dominate at low and high temperatures. At temperatures above \(\sim 30\) K correlated pinning by  sparser microstructures as insulating pins such as BaZrO$_3$ or RE$_2$O$_3$ dominates, while at lower temperatures, weak but dense pinning centers add equivalent and isotropic pinning \cite{Aixia2012}. Scaling \( I_c \) from 77 K to 4.2 K introduces an uncertainty of over 60\% \cite{Valeria2011}. Currently, no instrumentation can perform continuous lengthwise \( I_c(x) \) measurements at low temperatures.

Our results, obtained from short samples cut from long-length tapes, show significant  $I_c(x)$ variation within the same tape. Figure~\ref{fig9}a shows the angular \( I_c \) measured for several short samples cut every 20 m from the original tape. It is evident that the angular \( I_c \) measurements differ considerably near the \( ab \) plane. At \( \theta = 90^\circ \), differences can reach up to 50\%, whereas at \( \theta = 30^\circ \), the differences decrease to approximately 10\%. Figure~\ref{fig9}b shows \( I_c \) values of the four samples cut from another tape. Although these samples were cut at larger distances, \( I_c \) traces overlap with  accuracy of line thicknesses in Figure~\ref{fig9}b. The two tapes were manufactured with the same specifications; however, the homogeneous tape was cut from the center of a 12 mm wide production tape, whereas the tape with strong lengthwise variations was cut from the edge. The edge, which is not  centrally positioned during growth,  exhibits poor homogeneity. This observation is confirmed in three other tapes and  suggests that \( I_c \) variations stem from issues with growth near the edge. However, further studies are required to confirm this hypothesis.

\subsection{Flux jumps}
The same conditions that enable full current saturation also expose thermomagnetic instabilities.
During  torque measurements, a significant tendency for flux jumps is observed in certain REBCO conductors. 
Flux jumps in  type-II superconductors stem from the fact that the stationary penetrating field $B$ distribution depends on energy dissipating screening currents, with $I_c(B,T)$ strongly dependent on $B$ and $T$ and strong $I-V$ non-linearity. This leads to a fragile non-equilibrium state, and if the heat produced by the current is not effectively removed, thermomagnetic instability occurs. This instability leads to a sudden vortex redistribution and associated heating which can catastrophically drive the conductor normal \cite{Mints1996,Wilson1983}.

\begin{center}
\begin{figure}[htb]
\hspace{1pt}\hfill\includegraphics[width=.5\linewidth]{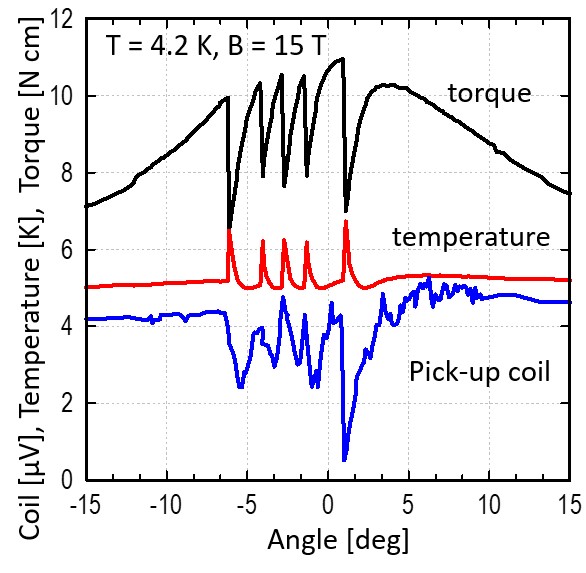}\hfill\hspace{1pt}\\
\caption{The torque, temperature and pick-up coil signal measured vs.\ increasing angle around the ab plane at $T=4.2$~K and at $B=15$~T. Sudden dropouts of the torque signal are accompanied by sharp temperature spikes and dropouts of the coil signal }\label{Fig10}
\end{figure}
\end{center}

Figure \ref{Fig10} illustrates the torque, temperature, and voltage from the pick-up coil measured as the sample is rotated in an external magnetic field  $ B = 15$~T at a speed of 0.1$^{\circ}$/s. Flux jumps manifest as sudden drops in the torque during sample rotation, accompanied by noticeable temperature increases of up to 7 K, as measured by a Cernox thermometer. The actual local temperature rise in the superconducting layer is expected to exceed the value measured by the thermometer. The flux jumps are also directly observed via the pick-up coil, whose primary purpose is to measure the sample angle, but it also clearly picks up the sample flux changes as well. The lowest trace in the figure shows a significant decrease in the signal from the pickup coil at the beginning of each flux jump, indicating a sudden reduction in the screening currents in the sample.

A systematic study of this phenomenon across various REBCO-coated conductors from different manufacturers led to several important conclusions, as  summarized in Table~1. Flux jumps occur more frequently in samples with thicker REBCO films. For instance, they are consistently observed in 5~$\mu$m thick samples at 4.2 K and even above 20 K. They occur in  3~$\mu$m thick samples at 4.2 K but not at 20 K. In contrast, 1~$\mu$m thick samples do not exhibit flux jumps at any $T$. Thicker samples generally carry higher currents, which increases losses and exacerbates thermal instability. Additionally, it is well known \cite{WERTHEIM1967,Bolz2003} that the quench velocity is inversely proportional to the sample dimensions. The frequency of flux jumps also increases with the rotation speed of the sample stage; they are present in  5~$\mu$m samples even at the slowest rotation rate (0.01$^{\circ}$/s). This suggests that manufacturing tapes with thick REBCO films may not be the best way to increase the entire tape current density, especially for devices operating at lower temperatures. For thinner samples, flux jumps occur only at faster rotation speeds, such as 1$^{\circ}$/s.

\begin{table}[t]
\caption{Conductors  showing flux jumps during rotation. Samples with thin REBCO layer do not show flux jumps at any rotation speed used (max.\ one $^{\circ}$/s).
These data strongly suggest the REBCO thickness correlates with flux jumps, while rotation speed and critical current also contribute. }
\begin{center}
\begin{tabular}{|p{0.12\textwidth}|p{0.37\textwidth}|p{0.1\textwidth}|p{0.12\textwidth}|r|}\hline
Samples&Flux jumps?&thickness [$\mu$m]&rotation speed $^{\circ}$/s & $I_c^{ab}$ [A] at 15 T\\\hline\hline
A (many)&always at 4.2~K and 20 K&5.0& 0.0125&6000 \\\hline\hline
B (many)&always at 4.2~K never at 20 K&$3.5$&0.125& 900\\\hline\hline
C&always at 4.2~K&$3.0$&0.0125& 1080\\\hline\hline
D (many)&often at 4.2~K&1.8&0.1& 2000\\\hline\hline
E (many)&sporadically at 4.2~K&1.8&0.5& 900\\\hline\hline
F&sporadically at 4.2~K&1.57&0.25& 2000\\\hline\hline
G&sporadically at 4.2~K&1.4&0.25& 1500\\\hline\hline
H&never&1.12&1&1800\\\hline\hline
\end{tabular}
\end{center}
\end{table}

In REBCO CC, flux jumps are more likely to be observed in high magnetic fields, in contrast to  low-temperature superconductors such as Nb$_3$Sn. Specifically,  critical current of REBCO decays very slowly with an increasing field around the \( ab \) plane: \( I_c \propto \exp(-B/B_0) \). With a typical \( B_0 \approx 40 \) T, \( I_c \) decreases slower than \( B \) increases, showing only a 40\% decrease between 5 and 30 T. Consequently, the Lorentz forces {\em increase} with a field up to 45 T, which is the highest field used, promoting instability. Improving the heat sink, such as  using a thicker copper stabilizer or measuring in liquid helium instead of helium exchange gas, helps suppress these jumps, highlighting the well known connection between flux jumps and thermal instability.

Rotating REBCO CC in a magnetic field leads to extreme conditions where the screening currents are fully saturated near the ab plane, where both $I_c$ and the external magnetic field flux change $d\Phi/dt$ reach their maximum values. So far, flux jumps have not been observed when the magnetic field is swept while the sample remains at a fixed angle, even at a maximum sweep rate of 10 T/min. 
This absence may be because the screening currents near the ab
 plane are significantly lower because they occupy only the edge of the conductor
 in a not fully penetrated state, as discussed earlier.
However, recent studies have shown flux jumps in REBCO stacks \cite{Tushar2025} when the perpendicular field is swept. Flux jumps remain a critical concern for magnet design, even if the tape  in the magnets is not explicitly rotated. Moreover, reports \cite{Dylan2021} indicate that despite reinforcements, REBCO tapes in magnets can experience rotation under strong torques induced by screening currents. If this rotation involves a stick-slip movement, it can cause sudden changes in the tape angle (large  $d\Phi/dt$) which may trigger flux jumps.

Although the thermomagnetic instability associated with flux jumps has been extensively researched for its theoretical interest, it remains an undesirable phenomenon from an application perspective. The effective management and mitigation of flux jumps are crucial for the stability and performance of superconducting devices. Although flux jumps have been studied in various superconductors, including crystalline YBCO \cite{Hope1999,Chabanenko2017}, reports specifically on REBCO-coated conductors are relatively rare \cite{Chabanenko1996,Bolz2003,Kovalev2021}. This rarity may stem from the belief that REBCO operates at temperatures significantly below 
 $T_c$, making flux jumps expected to be infrequent. However, a phenomenally high $I_c$ at low temperatures can trigger flux jumps  even when it is well below $T_c$.

The observed flux jumps at high fields indicate a potential thermal instability, which could limit the performance of superconducting magnets in real-world applications.
 Future efforts should focus on mitigating these instabilities through on one hand enhanced pinning mechanisms, and on   applying better conductor thermal stabilization.

\section{Summary}

This study demonstrates that torque magnetometry is a powerful tool for evaluating the performance of REBCO tapes, particularly under very high magnetic fields and near-ab-plane orientations. Approximately 300 different conductors have been assessed at  magnetic fields of up to 30 T and beyond. Our findings demonstrate that torque magnetometry is a robust technique for assessing critical current and  its lengthwise fluctuations, offering a path toward improving the design of superconducting magnets. REBCO tapes with different doping formulations exhibited distinct angular critical current behaviors, which are important for magnet design. Lengthwise variations in the critical current were observed, with tapes cut from the edges of production tapes showing larger variability than   those cut from the center. These variations are significantly stronger when $B$ is parallel to the  ab plane. The magnetometer identified flux jumps in   the samples with thick  REBCO layers, which are linked to thermal instabilities. This strongly suggests that, instead of increasing  thickness of REBCO and reducing the stabilizer thickness, an enhancement of pinning is preferable to increase the technical critical current density. In conclusion, this study  provides a new and efficient way to characterize 2G HTS tapes, offering insights that are critical for the development of high-field magnets and superconducting technologies.

\ack

We are   grateful for  fruitful discussions with Viktor Chabanenko, Alex Gurevich, Denis Markiewicz,  Mike Sumption, and Seungyong Hahn. 
Many thanks to our colleagues from the DC field facility at the MagLab for their excellent support: Clyde Martin, Robert Nowell, Troy Brumm, David Graf, and Ali Bangura.
The National High Magnetic Field Laboratory is supported by the National Science Foundation through NSF grants DMR-1644779 and DMR-2128556, as well as by the State of Florida. The magnetometer used in this study was developed as part of the User Collaboration Grant Program. Additional funding was provided by the DOE Office of Fusion Energy Sciences DOE-SC0022011.

\section*{References}
\bibliography{UCGP1}{}

\end{document}